\documentstyle[prd,aps,preprint,tighten]{revtex}

\begin{document}

\draft
\preprint{\begin{tabular}{r}
{\bf hep-ph/0011217} \\
{\bf LMU-00-15} \\
{~~}
\end{tabular}}

\title{Democratic Neutrino Mixing and Radiative Corrections}
\author{\bf Zhi-zhong Xing}
\address{Sektion Physik, Universit$\it\ddot{a}$t M$\it\ddot{u}$nchen,
Theresienstrasse 37A, 80333 M$\it\ddot{u}$nchen, Germany \\
and \\
Theory Division, Institute of High Energy Physics, 
P.O. Box 918, Beijing 100039, China \\
{(\it Electronic address: xing@theorie.physik.uni-muenchen.de)} }
\maketitle

\begin{abstract}
The renormalization effect on a specific ansatz of lepton mass 
matrices, arising naturally from the breaking of flavor democracy for 
charged leptons and that of mass degeneracy for light neutrinos, is
studied from a superhigh energy scale $M_0 \sim 10^{13}$ GeV to the 
electroweak scale in the framework of the minimal supersymmetric 
standard model. We find that the democratic neutrino mixing
pattern obtained from this ansatz may in general be instable against
radiative corrections. With the help of similar flavor symmetries 
we prescribe a slightly different scheme of lepton mass
matrices at the scale $M_0$, from which the democratic mixing
pattern of lepton flavors can be achieved, after radiative corrections,
at the experimentally accessible scales.
\end{abstract}

\pacs{PACS number(s): 14.60.Pq, 12.15.Ff, 12.60.-i, 11.10.Hi}

Recently a number of models of lepton mass matrices have been proposed 
at low energy scales \cite{Review}, at which their consequences on the
spectrum of neutrino masses and the mixing of lepton flavors 
can directly be confronted with the robust Super-Kamiokande data on
atmospheric and solar neutrino oscillations \cite{SK}. 
From the theoretical point of view, however,
a phenomenologically-favored scheme of lepton mass matrices
might only serve as the low-scale approximation of a more 
fundamental theory responsible for the lepton mass generation and
flavor mixing at superhigh energy scales. It is therefore desirable to
investigate the scale dependence of lepton mass matrices with the
help of the renormalization-group equations. So far some attempts 
have been made in this direction \cite{RC,FX99,RC2,Tanimoto}.

In this Brief Report we aim to study whether the democratic neutrino mixing 
pattern, which is indeed a nearly bi-maximal mixing 
pattern of lepton flavors arising from the breaking of flavor 
democracy for charged leptons and that of mass degeneracy for 
light neutrinos, can be stable or not against the effect of
radiative corrections from a superhigh energy scale $M_0 \sim 10^{13}$ GeV
to the electroweak scale $M_Z$ in the framework of the minimal supersymmetric
standard model. We find that the democratic neutrino
mixing pattern at $M_0$ is no longer of the same form at $M_Z$. But
it can still be obtained at low energy scales, if an additional term
preserving the symmetry of flavor democracy is introduced to the 
original neutrino mass matrix at the scale $M_0$. Therefore
the democratic neutrino mixing pattern at the experimentally
accessible scales might hint at certain lepton flavor symmetries 
at a superhigh scale, at which viable models of lepton mass
matrices can naturally be built.

Let us begin with a brief retrospection of the specific model of 
democratic neutrino mixing proposed first  
in Ref. \cite{FX96} at low energy scales.
The essential idea of this model is that the realistic 
textures of charged lepton and neutrino mass matrices might 
arise respectively from the breaking of 
$\rm S(3)_{\rm L}\times S(3)_{\rm R}$ and S(3) flavor 
symmetries:
\begin{eqnarray}
M_l & = & \frac{c^{~}_l}{3} \left (\matrix{
1       & 1     & 1 \cr
1       & 1     & 1 \cr
1       & 1     & 1 \cr} \right ) + \Delta M_l \; ,
\nonumber \\
M_\nu & = & c_\nu \left (\matrix{
1       & 0     & 0 \cr
0       & 1     & 0 \cr
0       & 0     & 1 \cr} \right ) + \Delta M_\nu \; ,
\end{eqnarray}
where $c^{~}_l$ and $c_\nu$ measure the
corresponding mass scales of charged leptons and neutrinos.
The explicit symmetry-breaking term $\Delta M_l$ 
is responsible for the generation of muon and electron masses, 
and $\Delta M_\nu$ is responsible for the breaking of neutrino
mass degeneracy. The lepton flavor mixing matrix 
results from the mismatch between the diagonalization of
$M_l$ and that of $M_\nu$, therefore its pattern depends 
crucially on the forms of $\Delta M_l$ and $\Delta M_\nu$.
It has been shown in Ref. \cite{FX96} that current data on
solar and atmospheric neutrino oscillations seem to favor
the following forms of $\Delta M_l$ and $\Delta M_\nu$:
\begin{eqnarray}
\Delta M_l & = & \frac{c^{~}_l}{3} \left ( \matrix{
-i\delta_l       & 0     & 0 \cr
0       & i\delta_l     & 0 \cr
0       & 0     & \varepsilon^{~}_l \cr } \right ) \; ,
\nonumber \\
\Delta M_\nu & = & c_\nu \left ( \matrix{
-\delta_\nu       & ~ 0     & 0 \cr
0       & ~ \delta_\nu     & 0 \cr
0       & ~ 0     & \varepsilon_\nu \cr } \right ) \; ,
\end{eqnarray}
where 
$(\delta_l, \varepsilon^{~}_l)$ and $(\delta_\nu, \varepsilon_\nu)$
are dimensionless perturbation parameters of small magnitude.
It is easy to obtain $m_\tau \approx c^{~}_l$, 
$m_\mu \approx 2|\varepsilon^{~}_l|m_\tau/9$, and
$m_e \approx |\delta_l|^2 m^2_\tau/(27m_\mu)$ in the lowest order 
approximation. As for neutrino masses, we have 
$m_1 = c_\nu (1 - \delta_\nu)$, $m_2 = c_\nu (1 + \delta_\nu)$,
and $m_3 = c_\nu (1 + \varepsilon_\nu)$.
The simultaneous diagonalization of
$M_l$ and $M_\nu$ leads to the lepton flavor mixing matrix $V$, 
which links the neutrino flavor eigenstates $(\nu_e, \nu_\mu, \nu_\tau)$
to the neutrino mass eigenstates $(\nu_1, \nu_2, \nu_3)$:
\begin{equation}
V = \left ( \matrix{
\frac{1}{\sqrt{2}} & \frac{-1}{\sqrt{2}} & 0 \cr
\frac{1}{\sqrt{6}} & \frac{1}{\sqrt{6}} & \frac{-2}{\sqrt{6}} \cr
\frac{1}{\sqrt{3}} & \frac{1}{\sqrt{3}} & \frac{1}{\sqrt{3}} \cr}
\right ) + \Delta V \; ,
\end{equation}
where 
\begin{equation}
\Delta V \; = \; i ~ \xi^{~}_V \sqrt{\frac{m_e}{m_\mu}}
~ + ~ \zeta^{~}_V \frac{m_\mu}{m_\tau} \; 
\end{equation}
holds in the next-to-leading order approximation with
\begin{eqnarray}
\xi^{~}_V & = &  \left ( \matrix{
\frac{1}{\sqrt{6}}      & ~ \frac{1}{\sqrt{6}} ~        &
\frac{-2}{\sqrt{6}} \cr
\frac{1}{\sqrt{2}}      & ~ \frac{-1}{\sqrt{2}} ~       & 0 \cr
0       & ~ 0 ~ & 0 \cr} \right ) \; ,
\nonumber \\
\zeta^{~}_V & = & \left ( \matrix{
0       & 0     & 0 \cr
\frac{1}{\sqrt{6}}      & \frac{1}{\sqrt{6}}    & \frac{1}{\sqrt{6}} \cr
\frac{-1}{\sqrt{12}}    & \frac{-1}{\sqrt{12}}  & \frac{1}{\sqrt{3}}
\cr} \right ) \; .
\end{eqnarray}
This is just a nearly bi-maximal lepton mixing pattern with large
CP violation \cite{FX96}. Neglecting the term $\Delta V$, which is 
remarkably suppressed 
by the small quantities $\sqrt{m_e/m_\mu} \approx 0.07$ and
$m_\mu/m_\tau \approx 0.06$ \cite{PDG}, one often refers to $V$ as the
democratic neutrino mixing pattern. 

Now we prescribe the same ansatz of lepton mass matrices, as that
introduced above, at superhigh energy scales. 
To be specific, we only consider the
simple possibility that the typical mass scale of light Majorana 
neutrinos is determined via the conventional seesaw mechanism \cite{SS} 
by the mass of a heavy right-handed neutrino 
$M_0 \sim 10^{13}$ GeV; namely $c_\nu \sim v^2/M_0$ \cite{Hall}, 
where $v$ is the electroweak vacuum expectation
value. The mass degeneracy of three active neutrinos is broken 
by the perturbative term $\Delta M_\nu$ in Eq. (1) at the scale $M_0$.
Below this typical scale, the neutrino mass matrix $M_\nu$ and the charged 
lepton mass matrix $M_l$ have quite simple running behaviors 
in the framework of the standard electroweak model or its 
minimal supersymmetric extension. The relevant renormalization-group 
equations, which describe the radiative corrections to lepton mass
matrices from the superhigh scale $M_0$ to the electroweak scale 
$M_Z$, have been derived by a number of authors 
in Refs. \cite{RC,FX99,RC2,Tanimoto}.
Subsequently we investigate whether the democratic neutrino
mixing pattern $V$ in Eq. (3) is stable or not against radiative corrections
in the framework of the minimal supersymmetric standard model (MSSM).

For simplicity we choose a specific flavor basis, in which the
lepton mass matrix $M_l$ is diagonal (namely, $M_l$ is transformed
into the diagonal 
form $\hat{M}_l = {\rm Diag}\{ m_e, m_\mu, m_\tau \}$) at the
scale $M_0$. In this basis and at the same scale, the 
corresponding neutrino mass matrix takes the form
$\hat{M}_\nu = V^* M_\nu V^\dagger$. Running $\hat{M}_l$ and
$\hat{M}_\nu$ down to the scale $M_Z$ by use of the 
renormalization-group equations in the framework of MSSM, 
one obtains the new lepton mass matrices $\hat{\bf M}_l$ and 
$\hat{\bf M}_\nu$. Obviously $\hat{\bf M}_l$ remains diagonal, 
but its mass eigenvalues are in general different from those of $\hat{M}_l$ 
due to radiative corrections \cite{RC,RC2}. At the scale $M_Z$, the form 
of $\hat{\bf M}_\nu$ reads explicitly as \cite{FX99}
\begin{equation}
\hat{\bf M}_\nu \; = \; \left (I_g I^6_t \right ) T_l \hat{M}_\nu T_l \; 
\end{equation}
with 
\begin{equation}
T_l \; = \; \left ( \matrix{
I_e & 0 & 0 \cr
0 & I_\mu & 0 \cr
0 & 0 & I_\tau \cr} \right ) \; ,
\end{equation}
in which $I_g$, $I_t$, and $I_\alpha$ (for $\alpha = e, \mu, \tau$) 
denote the corresponding evolution functions of the gauge couplings 
$g^{~}_1$ and $g^{~}_2$, the top-quark Yukawa coupling $f_t$, and the 
charged lepton Yukawa couplings $f_e$, $f_\mu$ and $f_\tau$:
\begin{eqnarray}
I_g & = & \exp \left [ +\frac{1}{16\pi^2}
\int^{\ln M_0}_{\ln M_Z} \left ( \frac{6}{5}g^2_1(\chi) +
6 g^2_2(\chi) \right ) {\rm d}\chi \right ] \;\; ,
\nonumber \\
I_t & = & \exp \left [ -\frac{1}{16\pi^2}
\int^{\ln M_0}_{\ln M_Z} f^2_t(\chi) {\rm d}\chi \right ] \;\; ,
\nonumber \\
I_\alpha & = & \exp \left [ -\frac{1}{16\pi^2}
\int^{\ln M_0}_{\ln M_Z} f^2_\alpha (\chi) {\rm d}\chi \right ] \;\; .
\end{eqnarray}
Note that the power of $I_t$ in the expression of $\hat{\bf M}_\nu$
depends on the definition of $I_t$ in Eq. (8). 
The overall factor $(I_g I^6_t)$ in Eq. (6) does not affect the 
relative magnitudes of the matrix elements of $\hat{M}_\nu$. 
Only the matrix $T_l$, which amounts to the unity matrix
at the energy scale $M_0$, can modify the texture of
the neutrino mass matrix from $M_0$ to $M_Z$.
The magnitude of $I_\tau$ may somehow deviate 
from unity, if $\tan\beta$ (the ratio of
Higgs vacuum expectation values in the MSSM) takes large values.
In contrast, $I_e \approx I_\mu \approx 1$ is 
an excellent approximation. Denoting
$\kappa \equiv I_e/I_\tau - 1 \approx I_\mu/I_\tau - 1$, one 
arrives from Eq. (8) at
\begin{equation}
\kappa  \approx  \frac{m^2_\tau}{16 \pi^2 v^2 \cos^2\beta }
\ln \frac{M_0}{M_Z} \; .
\end{equation}
It turns out that $\kappa \approx 0.03$
for $M_0 \sim 10^{13}$ GeV and $\tan\beta = 60$. 

With the help of Eqs. (3) and (6), one can straightforwardly
figure out the explicit expression of $\hat{\bf M}_\nu$ at 
the scale $M_Z$. To a good degree of accuracy, the small term 
$\Delta V$ of $V$ (namely, the small corrections from the ratios 
of charged lepton masses) is negligible in the calculation. 
We then make the transformations 
$V^\dagger \hat{\bf M}_l V \equiv {\bf M}_l$ and 
$V^{\rm T} \hat{\bf M}_\nu V \equiv {\bf M}_\nu$ at the scale $M_Z$.
Of course ${\bf M}_l$ should have a quasi-democratic texture
in the lowest order approximation, just as $M_l$.
The neutrino mass matrix ${\bf M}_\nu$ takes the following form:
\begin{equation}
{\bf M}_\nu \approx \left (I_g I^6_t I^2_\tau \right )
\left [ M_\nu +  2 \kappa 
\left (\Omega_\nu - \Lambda_\nu \right ) \right ] \; , 
\end{equation}
where the $\kappa$-induced term signifies the renormalization 
effect, and the constant matrices $\Omega_\nu$ and $\Lambda_\nu$
read as
\begin{eqnarray}
\Omega_\nu & \approx & c_\nu\left ( \matrix{
1 & 0 & 0 \cr
0 & 1 & 0 \cr
0 & 0 & 1 \cr} \right ) \; ,
\nonumber \\
\Lambda_\nu & \approx & \frac{c_\nu}{3} \left ( \matrix{
1 & 1 & 1 \cr
1 & 1 & 1 \cr
1 & 1 & 1 \cr} \right ) \; .
\end{eqnarray}
In obtaining Eqs. (10) and (11), we have neglected the small
contributions of ${\cal O}(\kappa^2)$, ${\cal O}(\kappa \varepsilon_\nu)$ 
and ${\cal O}(\kappa \delta_\nu)$. Note that the factor $I^2_\tau$ in 
${\bf M}_\nu$ comes from the product of two $T_l$ matrices on the right-hand 
side of $\hat{\bf M}_\nu$. Comparing ${\bf M}_\nu$ with
$M_\nu$, we see that the diagonal texture of $M_\nu$ is not 
affected by the radiative correction term $\Omega_\nu$, which is 
also diagonal. The latter modifies three neutrino mass eigenvalues 
of $M_\nu$ with the same magnitude (proportional to $2\kappa c_\nu$).
In contrast, the diagonal texture of $M_\nu$ is spoiled by the 
other radiative correction term $\Lambda_\nu$, which has a democratic 
form. As a consequence of the appearance of $\Lambda_\nu$ in
${\bf M}_\nu$, the corresponding lepton flavor mixing matrix $\bf V$, which 
arises from the mismatch between the diagonalization of ${\bf M}_l$ and 
that of ${\bf M}_\nu$ at the electroweak scale $M_Z$, may substantially 
deviate from the original flavor mixing matrix $V$ at the superhigh scale 
$M_0$. Unless $\kappa$ is negligibly small, the democratic neutrino mixing 
pattern is expected to be instable against radiative corrections.

The sensitivity of a nearly bi-maximal neutrino mixing pattern 
to the renormalization effect is of course not a big surprise \cite{RC,RC2}.
However, it is not impossible to find out the appropriate textures of 
lepton mass matrices, which are essentially stable against radiative
corrections \cite{RC2,Tanimoto}. As the democratic neutrino mixing 
pattern is only favored at the experimentally accessible energy scales,
the question turns out to be whether there is a scheme of lepton mass
matrices at the superhigh scale $M_0$, from which the democratic
mixing pattern of lepton flavors can be obtained at the electroweak
scale $M_Z$. We find that such a scheme of lepton mass matrices does
exist and it is very similar to that discussed above. 

To be specific, let us prescribe the new ansatz of lepton mass 
matrices at the scale $M_0$. We take the charged lepton
mass matrix $M'_l$ to have the same form as $M_l$ in Eq. (1);
namely, $M'_l = M_l$ has the $\rm S(3)_{\rm L}\times S(3)_{\rm R}$
flavor symmetry in the limit $\Delta M_l = 0$. 
The corresponding neutrino mass matrix $M'_\nu$
is a linear combination of $M_\nu$ in Eq. (1) and 
an additional term, which preserves the symmetry of flavor 
democracy:
\begin{equation}
M'_\nu \; =\; M_\nu + 
c'_\nu \left ( \matrix{
1 & 1 & 1 \cr
1 & 1 & 1 \cr
1 & 1 & 1 \cr} \right ) \; .
\end{equation}
Obviously $M'_\nu$ has the same S(3) flavor symmetry as $M_\nu$ in
the limit $\Delta M_\nu =0$. One can therefore see much similarity
between the new ansatz and the old one. The coefficient $c'_\nu$
is a free parameter \cite{Tanimoto}, but its value may be physically 
nontrivial, as we shall see below.

Following the same procedure as outlined above, we calculate the
counterpart of $M'_\nu$ at the electroweak scale $M_Z$. We obtain
\begin{equation}
{\bf M}'_\nu \approx \left (I_g I^6_t I^2_\tau \right )
\left [ M'_\nu +  2 \kappa 
\left (\Omega_\nu - \Lambda_\nu \right ) \right ] \; ,
\end{equation}
where $\kappa$, $\Omega_\nu$, and $\Lambda_\nu$ have been given 
in Eqs. (9) and (11). Note that the texture of $\Omega_\nu$ is
essentially the same as that of $M_\nu$ (the first term of 
$M'_\nu$), and the texture of $\Lambda_\nu$ is essentially the same 
as that of the second term of $M'_\nu$. Therefore the basic texture
of ${\bf M}'_\nu$ is the same as that of $M'_\nu$. In other words,
the structure of $M'_\nu$ is stable against radiative corrections.

It is particularly interesting that the second term of $M'_\nu$ 
and the term $\Lambda_\nu$ in ${\bf M}'_\nu$ are possible to cancel
eath other. Indeed the cancellation between these two terms
takes place, if the coefficient $c'_\nu$ satisfies the
condition $c'_\nu/c_\nu = 2 \kappa /3$. 
In this case, the resultant neutrino mass matrix reads as
\begin{equation}
{\bf M}''_\nu \approx \left (I_g I^6_t I^2_\tau \right )
\left ( M_\nu +  2 \kappa \Omega_\nu \right ) \; ,
\end{equation}
which is diagonal at the scale $M_Z$. 
Since the corresponding charged lepton mass matrix ${\bf M}'_l$
is of a quasi-democratic form at the same scale, just as ${\bf M}_l$
discussed above, the simultaneous diagonalization of ${\bf M}'_l$
and ${\bf M}''_\nu$ must lead to the democratic flavor mixing
pattern $V$ in the lowest order approximation (namely, in the
approximation of neglecting the $\Delta V$ term). 

One might question whether the condition $c'_\nu/c_\nu = 2 \kappa /3 \ll 1$ 
suffers from fine-tuning or not. 
Indeed it is rather natural to expect $c'_\nu \ll c_\nu$ in the
model under consideration, because this hierarchy assures the near
degeneracy of three neutrino masses. On the other hand, a relation
between $c'_\nu$ and $c_\nu$ allows us to reduce the number of free
parameters in $M'_\nu$ from four to three, which can fully be determined
by three neutrino mass eigenvalues ${\bf m}'_i$ (for $i=1,2,3$) at
low energy scales. Indeed we obtain 
\begin{eqnarray}
{\bf m}'_1 & \approx & \left (I_g I^6_t I^2_\tau \right )
\left (1 + 2\kappa - \delta_\nu \right ) c_\nu \; ,
\nonumber \\
{\bf m}'_2 & \approx & \left (I_g I^6_t I^2_\tau \right )
\left (1 + 2\kappa + \delta_\nu \right ) c_\nu \; ,
\nonumber \\
{\bf m}'_3 & \approx & \left (I_g I^6_t I^2_\tau \right )
\left (1 + 2\kappa + \varepsilon_\nu \right ) c_\nu \; .
\end{eqnarray}
Note that the overall factor $(I_g I^6_t I^2_\tau)$ takes approximate
values 0.80 and 0.63, respectively, for $\tan\beta = 10$ and $60$ \cite{FX99}. 
The present Super-Kamiokande \cite{SK} and CHOOZ \cite{CHOOZ} data
favor the approximate decoupling between solar and atmospheric neutrino 
oscillations, which are respectively attributed to 
$\nu_e \rightarrow \nu_\mu$ and 
$\nu_\mu \rightarrow \nu_\tau$ transitions in the framework of three
active neutrinos.Thus one may take
$\Delta m^2_{\rm sun} = | ({\bf m}'_2)^2 - ({\bf m}'_1)^2|$ and
$\Delta m^2_{\rm atm} = | ({\bf m}'_3)^2 - ({\bf m}'_1)^2|$.    
Taking $\Delta m^2_{\rm sun} \ll \Delta m^2_{\rm atm}$ into 
account \cite{SK}, we arrive at
\begin{equation}
\frac{\Delta m^2_{\rm sun}}{\Delta m^2_{\rm atm}} \; \approx \;
\frac{2|\delta_\nu|}{|\varepsilon_\nu + \delta_\nu|} \; \approx \;
2 \frac{|\delta_\nu|}{|\varepsilon_\nu|} \;\; .
\end{equation}
This result implies that the observables $\Delta m^2_{\rm sun}$ and 
$\Delta m^2_{\rm atm}$ are completely insensitive to the small
parameter $\kappa$. Given the energy scale at 
which the proposed textures of $M'_\l$ and $M'_\nu$ hold, $\kappa$ is a
well-defined quantity in respect to the fixed value of $\tan\beta$ 
within the MSSM or other extensions of the standard electroweak model. 
Therefore we think that the condition $c'_\nu/c_\nu = 2 \kappa /3$ is 
plausible for our new ansatz of lepton mass matrices at a superhigh 
energy scale, from which the democratic neutrino mixing pattern can 
be obtained, after radiative corrections, at
the experimentally accessible energy scales. 

In summary, we have investigated the renormalization
effects on lepton mass matrices and flavor mixing from a superhigh
energy scale to the electroweak scale in the framework of MSSM. 
We find that the
democratic neutrino mixing pattern may in general be instable against
radiative corrections. A new ansatz of lepton mass matrices, based
on the breaking of flavor democracy for charged leptons and the
mass degeneracy for light neutrinos, has been prescribed
at superhigh scales. Taken the effect of radiative corrections 
into account, this ansatz can lead to the democratic
mixing pattern of lepton flavors at low energy scales. We expect that
the forthcoming neutrino oscillation experiments will provide a
stringent test of the democratic neutrino mixing pattern and other
nearly bi-maximal neutrino mixing schemes, from
which one can get more hints to explore possible flavor symmetries
and to build viable models of lepton mass matrices at appropriate
superhigh energy scales.

The author would like to thank H. Fritzsch, N. Haba, and M. Tanimoto
for useful discussions. He is also grateful to K.R.S. Balaji for
helpful comments.

\end{document}